\documentclass[final,5p,times,twocolumn]{elsarticle}
\usepackage{graphicx}
\usepackage{amsmath}
\usepackage{geometry}
\usepackage{amssymb}

\journal{One Journal}

\begin{document}
	
\begin{frontmatter}

\title{Plano-concave microlenses with epsilon-near-zero surface-relief coatings \\ for efficient shaping of nonparaxial optical beams}
%\title{Generation of nonparaxial focal waves and accelerating beams using \\ subwavelength-structured epsilon-near-zero materials}

\author[label1,label2]{Mahin Naserpour}
\author[label1]{Carlos J. Zapata-Rodr\'{i}guez\corref{cor*}}
\cortext[cor*]{Corresponding author:}
\ead{carlos.zapata@uv.es}
\author[label3]{Mahdieh Hashemi}
\address[label1]{Department of Optics and Optometry and Vision Science, University of Valencia, Dr. Moliner 50, Burjassot 46100, Spain}
\address[label2]{Department of Physics, College of Sciences, Shiraz University, Shiraz 71946-84795, Iran}
\address[label3]{Department of Physics, College of Science, Fasa University, Fasa 7461781189, Iran}

\begin{abstract}
Epsilon-near-zero (ENZ) materials, including artificial metamaterials, have been advanced to mold laser beams and antenna-mediated radiated waves.
Here we propose an efficient method to control Ohmic losses inherent to natural ENZ materials by the assembly of subwavelength structures in a nonperiodic matrix constituting an ENZ metacoating.
Implemented over plano-concave transparent substrates whose radius can be of only a few wavelengths, ENZ surface-relief elements demonstrate to adequately shape a plane wave into highly localized fields.
Furthermore, our proposal provides an energy efficiency even higher than an ideally-lossless all-ENZ plano-concave lens.
Our procedure is satisfactory to generate aberration-free nonparaxial focused waves and accelerating beams in miniaturized spaces.
\end{abstract}


\end{frontmatter}

\section{Introduction}

Dielectric microlenses used for imaging and focusing in miniaturized optoelectronic devices are currently prospects to be substituted by metalenses and metasurfaces exhibiting an extraordinary optical performance within notably reduced volumes.
Such engineered materials include elementary scatterers such as nanoslits \cite{Verslegers09,Ishii11,Naserpour16} and nanoholes \cite{Lin10,Matsui12,Ishii13}, split-ring resonators \cite{Wang15b,Hashemi16}, and V-shaped nanoantennas \cite{Aieta12b,Yu13}, to mention a few, enabling the simultaneous control of the phase front and polarization of an incoming plane wave.
In fact, beam shaping based on holographic metamaterials may be further applied to a general class of wave fields \cite{Genevet15}.

In particular, epsilon-near-zero (ENZ) materials have been proposed theoretically \cite{Alu07,Navarro12,Memarian15} and demonstrated experimentally \cite{Pacheco14,Pacheco16} for far-field light focusing due to its enhanced wave directivity experienced at boundaries with dielectric environments.
Since no phase delay occurs inside the ENZ medium, one can manipulate the phase radiation pattern of a given impinging phase front and transform it into a desired shape by properly sculping the exit surface of the ENZ slab. 
For photonic applications, however, the intrinsic losses of natural Drude materials at its plasma frequency and of artificial materials designed to undergo a zero effective permittivity severely restricts their use at propagation depths of several wavelengths \cite{Caldwell15}. 
Even using all-dielectric zero-index photonic crystals, the lens performance can be notably undermined due to radiative losses, interface losses, and imperfections in the nanofabrication \cite{Casse08,He16}.
Zoned plates are then good candidates with an increased efficiency provided that the slab thickness remains in the scale of the wavelength \cite{Pacheco13,Orazbayev15}. 
Nevertheless, their planar structure circumscribes the focal architecture to low and moderate numerical apertures thus restricting the achievable resolution power of the optical device.

Here we propose the use of epsilon-near-zero materials adhered to the terraced exit surface of plano-concave transparent substrates to efficiently shape a plane wave into an aberration-free nonparaxial focused wave.
Our procedure is also adequate to generate nonparaxial accelerating beams with Bessel signature in miniaturized spaces \cite{Zapata14c}.
We design surface-relief structures using both metals and doped semiconductors as ENZ materials spanning the mid-IR, visible and UV spectral bands.

\section{ENZ plano-concave lenses to focus and accelerate optical beams}

\begin{figure}[tb]
	\centering
	\includegraphics[width=\linewidth]{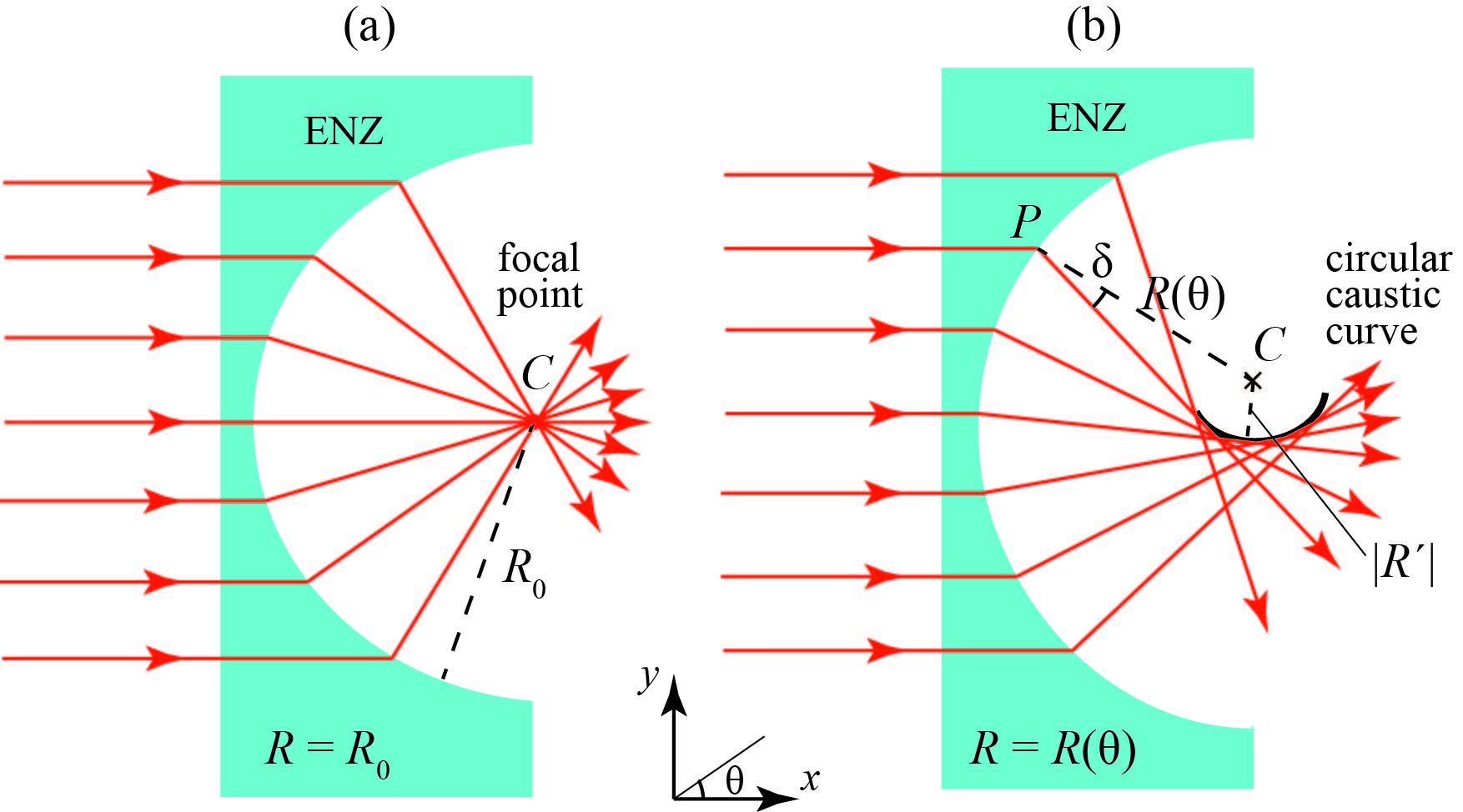}
	\caption{Plano-concave focusing devices of high numerical aperture made of ENZ materials.
		(a) Perfectly spherical exit surface enables to create aberration-free nonparaxial focal fields.
		(b) Angular-resolved curvature to generate concentric caustic fields.}
	\label{fig01}
\end{figure}

A plano-concave ENZ lens can transform a plane wave into a cylindrical converging wave field, as depicted in Fig.~\ref{fig01}(a).
Here we adopted a two-dimensional geometry where the concave surface has an axis of revolution at $C$; note that a converging spherical wavefront can be obtained by using concave surfaces whose center of symmetry is set at the focal point.
A simple interpretation can be followed by using optical rays.
When a collimated beam impinges normally to the front plane surface of the ENZ lens, the optical rays does not undergo any deviation in its trajectory.
On the other hand, light emerges from an ENZ material forming a zero angle of refraction with respect to its normal line.
This can be straightforwardly deduced by applying the Snell's law over the exit surface of the plano-concave lens.
Therefore, an ENZ material bounded by a surface with center of curvature at the point $C$ will refract light in direction to such a point.
Small deviations of the curvature will produce aberrated focal fields that, in general, are undesirable.

For simplicity, let us consider light rays propagating in the incidence plane, that is the $xy$ plane, which is perpendicular to the axis of the cylindrical concave surface of the lens.
If the function $R = R\left( \theta \right)$ represents the curve at the exit surface in polar coordinates, as drawn within the incidence plane, the trajectory of a light ray is driven by the normal vector to such curve, $N = - \hat r + \hat \theta R' / R$, where $\hat r$ and $\hat \theta$ are unitary vectors and $R' = \partial R / \partial \theta$ is a first derivative evaluated at the point of interest over the curve.
Defining $\delta$ as the angle of the trajectory of the light ray with respect to the unitary vector $- \hat r$, the latter giving the ray trajectory in the case of a perfect cylindrical surface as illustrated in Fig.~\ref{fig01}(b), we finally obtain that
\begin{equation}
 \tan \delta = \frac{|R'|}{R} .
\end{equation}
Positive values of $R'$ will make the light rays to pass below the point $C$ thus forming a shadow region.

Controlled curvature engineering can give, as a result, patterned fields near the focal point.
As an example illustrated in Fig.~\ref{fig01}(b), let us consider an exit surface with azimuthally-varying radius of curvature $R$ given by
\begin{equation}
 R \left( \theta \right) = R_0 + R' \left( \theta - \pi / 2 \right) ,
 \label{eq01}
\end{equation}
where $R_0$ stands for the initial radius at $\theta = \pi / 2$ as measured from the curvature center $C$ and $R'$ is here a constant term indicating the rate of angular variation of the surface radius.
As noted above, such curvature deviation of the exit surface leads to a prismatic effect on all the emerging light rays.
Let us first assume a modest deviation of the perfectly circular symmetry so that $|R'| \ll R_0$.
Now it is clear that a ray emerging from a given point $P$ of the exit surface will undergo an angular deviation $\delta = |R'| / R_0$ with respect to the segment connecting $P$ and $C$.
As a consequence, the rays propagating in the $xy$-plane will generate a caustic curve around the focal point $C$, which is circular with center also at $C$ and with a radius of curvature given by $|R'|$.

\begin{figure}[tb]
	\centering
	\includegraphics[width=\linewidth]{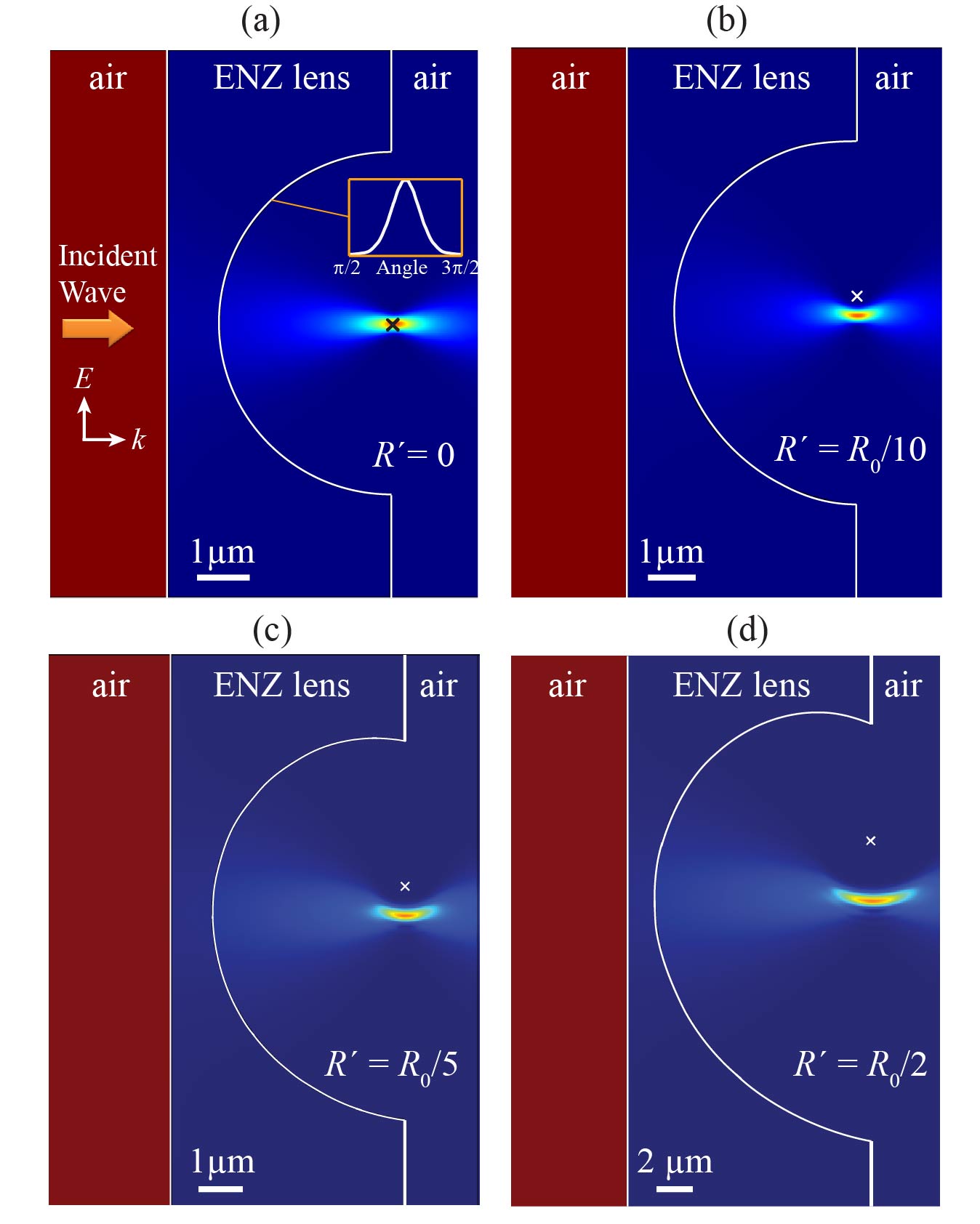}
	\caption{Time-averaged energy density generated by a plano-concave lossless ENZ lens immersed in air, provided that the incident plane wave is polarized along the $y$ axis, that is TM-polarized, at a wavelength $\lambda = 326$~nm.
		The exit surface is (a) perfectly circular with $R' = 0$, where the symbol {\tiny $\times$} denotes the center point $C$; and it is shaped by the function given in Eq.~(\ref{eq01}) with (b) $R' = R_0 / 10$, (c) $R' = R_0 / 5$, and (d) $R' = R_0 / 2$.
		In all cases $R_0 = 10 \lambda$.
		Inset in (a): Apodizing time-averaged energy density along the concave exit surface.}
	\label{fig02}
\end{figure}

An analogous interpretation can be given in terms of wave optics.
The phase shift accounted in the optical path of rays traveling from the concave surface to the focus $C$ is given by
\begin{equation}
 \exp \left( -i k R \right) = \exp \left( -i k R_0 \right) \exp \left( -i k R' \theta \right) ,
\end{equation}
where $k = 2 \pi / \lambda$ is the wavenumber in vacuum.
The term $\left( -i k R_0 \right)$ is related with the phase shift produced by an aberration-free converging lens.
Assuming that $R' \ll R_0$, $\exp \left( -i k R' \theta \right)$ can be set as a linear phase shift $\exp \left( - i k l R'/R_0 \right)$, where $l$ represents an spatial coordinate as the arc length measured along the reference surface of radius $R_0$.
Thus the wavefront emerging from the vicinities of a point $P$ of the exit surface will cross the reference surface obliquely, undergoing an angular deviation $\delta = - R' / R_0$.
This is in agreement with the fact that $\delta$ also denotes the angular deviation of the normal line to the exit surface with respect to the segment $\overline{PC}$.
If finally we include the given term of the form $\exp \left( i m \theta \right)$, denoting the field in the reference surface, in a diffraction integral (for instance the Debye diffraction integral \cite{Zapata07}) to estimate the field distribution near the focal point $C$, where in our case the parameter $m = - k R'$, it might be demonstrated that the amplitude of the diffracted field results in direct proportion to $\exp \left(i m \theta \right) J_m \left( k r \right)$ near the caustic curve, where $r$ is the radial coordinate and $J_m \left( \cdot \right)$ is the Bessel function of the first kind and order $m$; a full demonstration can be found in Ref.~\cite{Naserpour15b}. 
Such wave function has an intensity with maximum around $C$ at a distance approximately given by $r_m = |m|/k$, that is $|R'|$.
The latter is consistent with our previous description given in terms of light rays.

To illustrate the capabilities of a plano-concave ENZ lens to focus and accelerate an incident plane wave, we numerically estimate the scattered wave field by means of a commercial full-wave solver of the Maxwell's equations (COMSOL Multiphysics) based on the finite element method (FEM).
In Fig.~\ref{fig02}(a) we show the spatial distribution of the time-averaged energy density when a plane wave that is linearly polarized along the $y$ axis impinges normally on the flat entrance surface of the ENZ lens.
We consider a wavelength $\lambda = 326$~nm for which the dielectric constant of silver has its real part dropping to zero.
In this case, the concave surface is perfectly circular, that is $R' = 0$, with a radius $R_0 = 10 \lambda$.
For numerical purposes, we first consider a slightly lossy ENZ material with relative permittivity $\epsilon = 0 + i 0.001$.
In the figure we observe the formation of a light spot centered at $C$ with a FWHM of $184$~nm in the transverse direction, that is the $y$ axis, and a width of $504$~nm along the optical axis.
This means that the effective numerical aperture (NA) is approximately $0.90$, in agreement with the fact that the effective angular semi-aperture of the plano-concave lens is $64^\circ$ \cite{Naserpour15d}.
Such decrement of the angular semi-aperture from the optimal $\pi / 2$ radians, as shown in the inset of Fig.~\ref{fig01}(a), should be attributed to a significant reduction of the lens transmittance at angles $|\theta| \approx \pi / 2$.
As a consequence, focal waves undergo an apodizing effect where sidelobes are notably reduced, however, remaining its central peak with an slightly greater width than potentially expected.

In Fig.~\ref{fig02}(b) we modulate the radius $R(\theta)$ of the lens exit surface such that $R' = R_0 / 10$.
As discussed above, we may form an accelerating beam with circular trajectory around a circumference of radius $\sim \lambda$.
Note that the acceleration of the optical beam can be produced above the point $C$ provided that $R' < 0$.
However, the trajectory of the beam is shorter than expected; note that a caustic curve with angular distribution of $\pi$ radians would hypothetically be formed, as inferred from the geometric picture given in Fig.~\ref{fig01}(b).
The decrease of the effective angular aperture of the plano-concave lens is obviously affecting to the trajectory length of the accelerating beam \cite{Naserpour15b}.

We point out that the acceleration of the focused field is also produced even when the condition $R' \ll R_0$ is not satisfied.
In Fig.~\ref{fig02}(c) and (d) we represent the intensity distribution of the magnetic field scattered by a plano-concave lens with high values of the parameter (c) $R' = R_0 / 5$ and (d) $R' = R_0 / 2$.
From our FEM-based numerical simulations, the radius of the caustic curve for the first case is approximately $2.0 \lambda$, whereas its radius increases to $5.1 \lambda$ for the case (b), which roughly are equal to $|R'|$.
These examples demonstrate the effective implementation of the required manipulation of the phase front.

Finally we point out that the polarization of the incident plane wave is reflected in a neglecting deviation of the spatial distribution of the focal waves shown in Figs.~\ref{fig02}(a)-(d).
However, a nonvanishing imaginary part of the permittivity of the ENZ material will make a difference as discussed below.

\subsection{Application to natural ENZ materials}

\begin{figure}[tb]
	\centering
	\includegraphics[width=\linewidth]{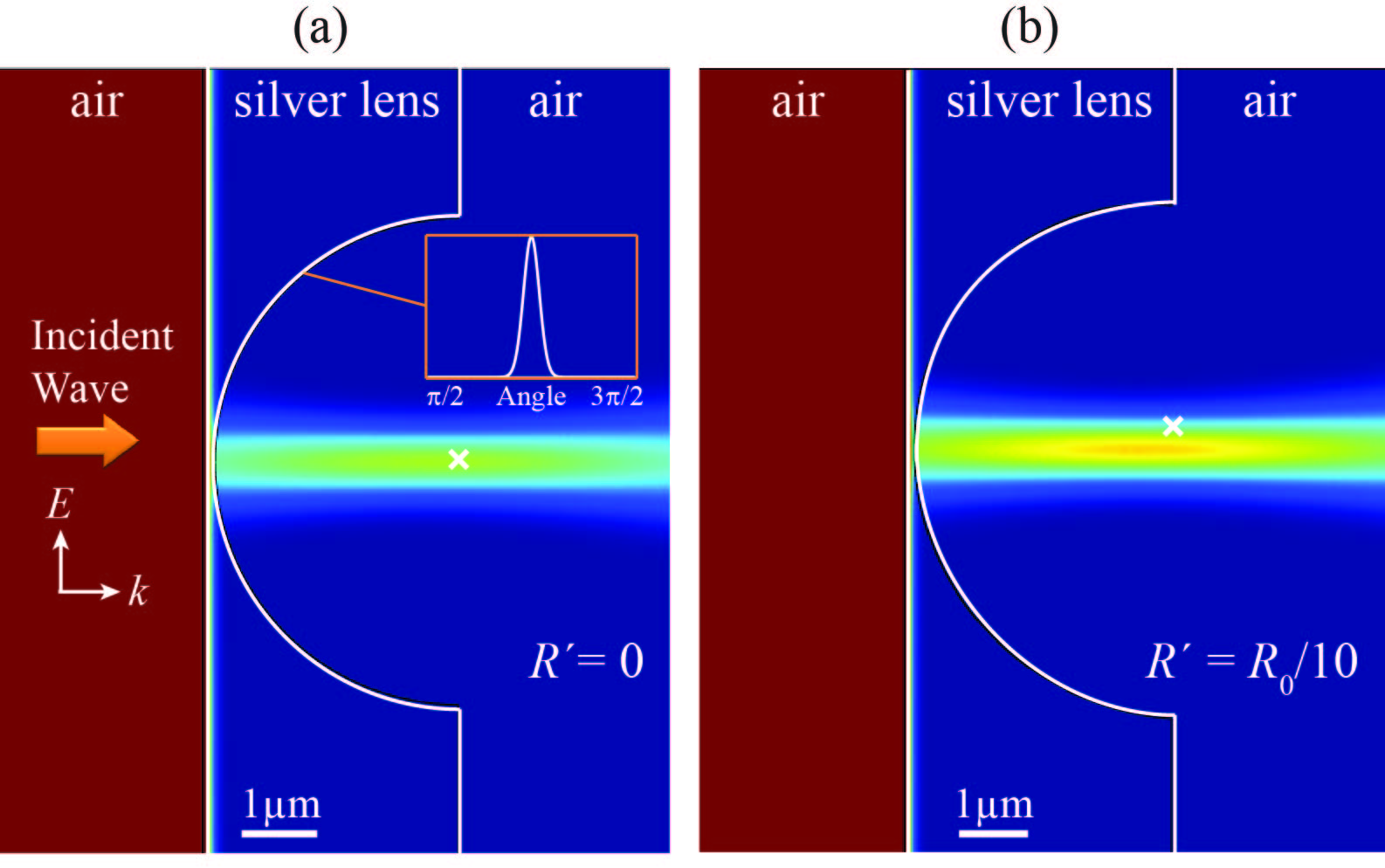}
	\caption{Focal distribution of the time-averaged energy density produced by a plano-concave silver lens of radius $R_0 = 10 \lambda$ at $\lambda= 326$~nm.
		The electric field of the incident plane wave is oriented along the $y$ axis.
		In (a) there is no modulation of the radius, $R' = 0$, and in (b) the modulation is determined by $R' = R_0 / 10$.
		Color scaling in the focal region refers to an energy range lower than in Fig.~\ref{fig02} to better depict the field pattern.}
	\label{fig03}
\end{figure}

Let us apply the previous concepts of beam shaping to a realistic ENZ material.
For that purpose we consider silver at a wavelength of $\lambda= 326$~nm, for which its permittivity has a vanishing real part.
However, the imaginary part of the relative permittivity of silver cannot be neglected, which is experimentally estimated as $\mathrm{Im} (\epsilon) = 0.7$ \cite{Caldwell15}. 
In Fig.~\ref{fig03}(a) we show the profile of the time-averaged energy density of the electromagnetic field focused by a plano-concave silver lens whose radius of curvature is $R_0 = 10 \lambda$.
In comparison with Fig.~\ref{fig02}(a), we first observe a clear decrease of the transmitted field, achieving a reduction of $58 \%$ at the focal point, a fact that is attributed to absorption in the lossy metal.
In fact, the maximum concentration of energy density is not located at $C$ but displaced towards the lens, an effect that is commonly coined as focal shift \cite{Zapata00}, in agreement with the fact that the effective numerical aperture of the lens has been reduced to $0.085$ (semi-aperture angle of only $\sim 5 ^\circ$), as illustrated in the inset of Fig.~\ref{fig03}(a).
On the other hand, the resolution power has dropped considerably; the FWHM in the transverse direction is now $751$~nm.
We might conclude that the limit of resolution is not determined by the angular aperture of the concave exit surface, but the metallic losses which are evidenced in the thick lens.

In Fig.~\ref{fig03}(b) we analyze the scattered field produced by a plano-concave lens with an angular-resolved curvature given by $R' = R_0/10$.
Essentially, we cannot recognize a different response in comparison with the case shown in Fig.~\ref{fig03}(a).
The expected acceleration of the field is fundamentally carried out at off-axis angles $\theta \neq \pi$ where the lens presents a higher thickness, and therefore dissipative effects are largely manifested.
Although not shown in the figures, higher values of $R'$ indicating faster modulation of the surface curvature provide barely a similar result.

In the FEM-based numerical simulations shown in Fig.~\ref{fig03}(a) and (b) we considered TM-polarized plane waves, that is the electric field is oriented along the $y$ axis, which are impinging over the plano-concave metallic lens.
TE-polarized plane waves particularly provide inferior focalization and acceleration due to a lower transmittance.
Such a distinct behaviour is ultimately caused by the nonvanishing imaginary part of the permittivity characterizing the ENZ material.

Finally, alternate ENZ materials like noble metals at the plasma frequency (and anisotropic metamaterials) can be analyzed for the use of plano-concave lenses to generate nonparaxial focused waves and accelerating beams at different wavelengths.
However, the inherent lossy characteristics of them makes the targeted beam shaping impracticable, even for a lens thickness of a few wavelengths.
For that reason, it is preferably to use alternate architectures including ultrathin ENZ layers to actively mold the wavefront of the incident plane wave.

\section{Implementation of ultrathin ENZ metacoatings}

\begin{figure}[tb]
	\centering
	\includegraphics[width=.5\linewidth]{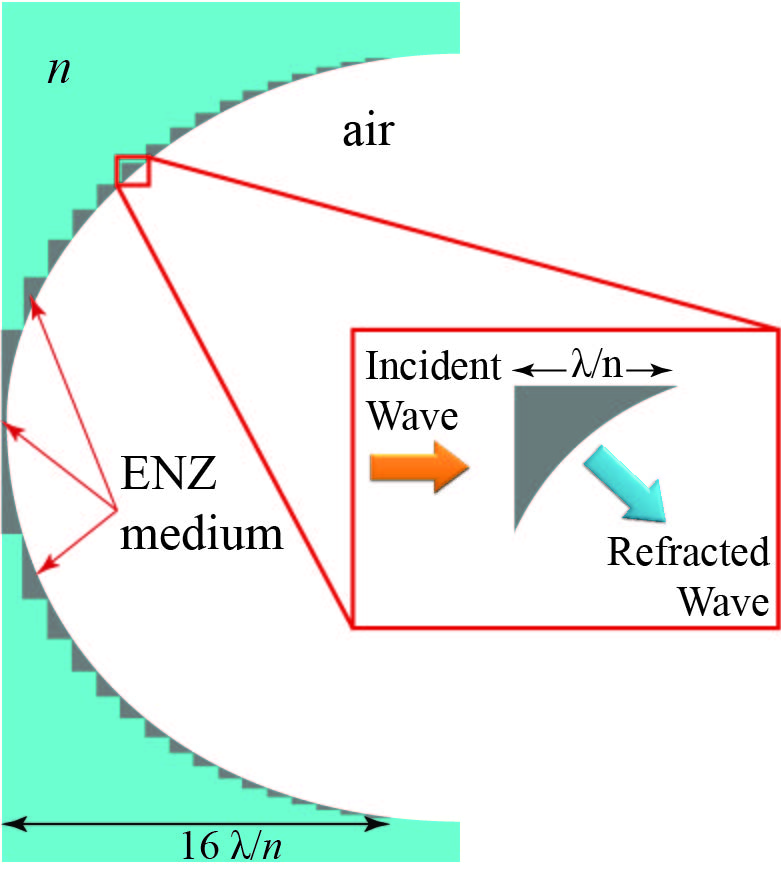}
	\caption{Schematic representation of the multi-prism ENZ nanostructure coating the terraced-concave surface of the dielectric thick lens, whose index of refraction is $n$. 
		Inset: Zoom-in view around a single ENZ subwavelength element composing the ultrathin coating.
		Here, the incident wave field interacts at the entrance (vertical) surface and the beam emerges from the exit curved surface.}
	\label{fig04}
\end{figure}

We propose to substitute the ENZ plano-concave lens by a lossless dielectric plano-concave lens, which inherently is divergent, also including a surface-relief ENZ ultrathin coating.
The latter will perform the necessary steering of the incident plane wave to generate either the focused field or the accelerating beam.
Since in addition the thick dielectric lens will not absorb the electromagnetic energy of the impinging wave field, we are able to optimally maintain the effective numerical aperture of the focusing architecture. 
As depicted in Fig.~\ref{fig04}, the ENZ elements of the coating have a straight (vertical) side where the plane wave is coupled by normal incidence, whereas they also exhibit a designed curved side where the beam emerges to be focused and accelerated, thus producing a prismatic effect.
Furthermore, the width of each ENZ element is given by $\lambda / n$, where $n$ denotes the refractive index of the plano-concave dielectric lens.
In this case, the incident field will undergo a phase shift of $2 \pi$ radians when it is coupled in adjacent ENZ elements, enabling in-phase interference of the divided incident wavefront within the focal region.
Finally, a reduced Ohmic loss in the transit of the wave field through each ENZ element is supported by its subwavelength width. 

\begin{figure}[tb]
	\centering
	\includegraphics[width=\linewidth]{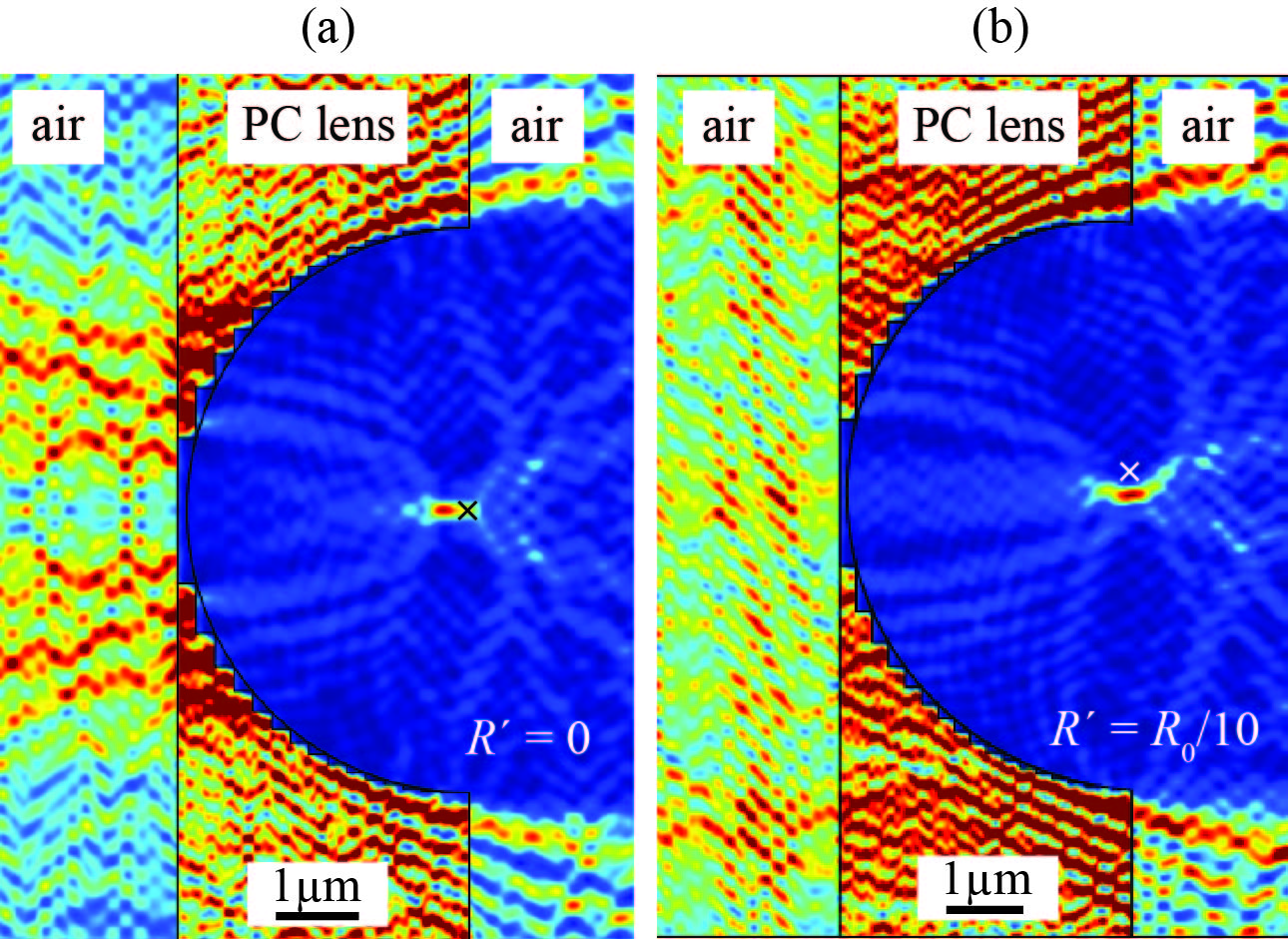}
	\caption{Time-averaged energy density generated by a plano-concave (PC) lens with surface-relief silver coating sculptured with an angular-resolved curvature $R_0 = 10 \lambda$ and (a) $R' = 0$, (b) $R' = R_0 / 10$.
		Again, the incident plane wave is TM-polarized at a wavelength $\lambda = 326$~nm.}
	\label{fig05}
\end{figure}

\begin{figure*}[h!]
	\centering
	\includegraphics[width=0.8\linewidth]{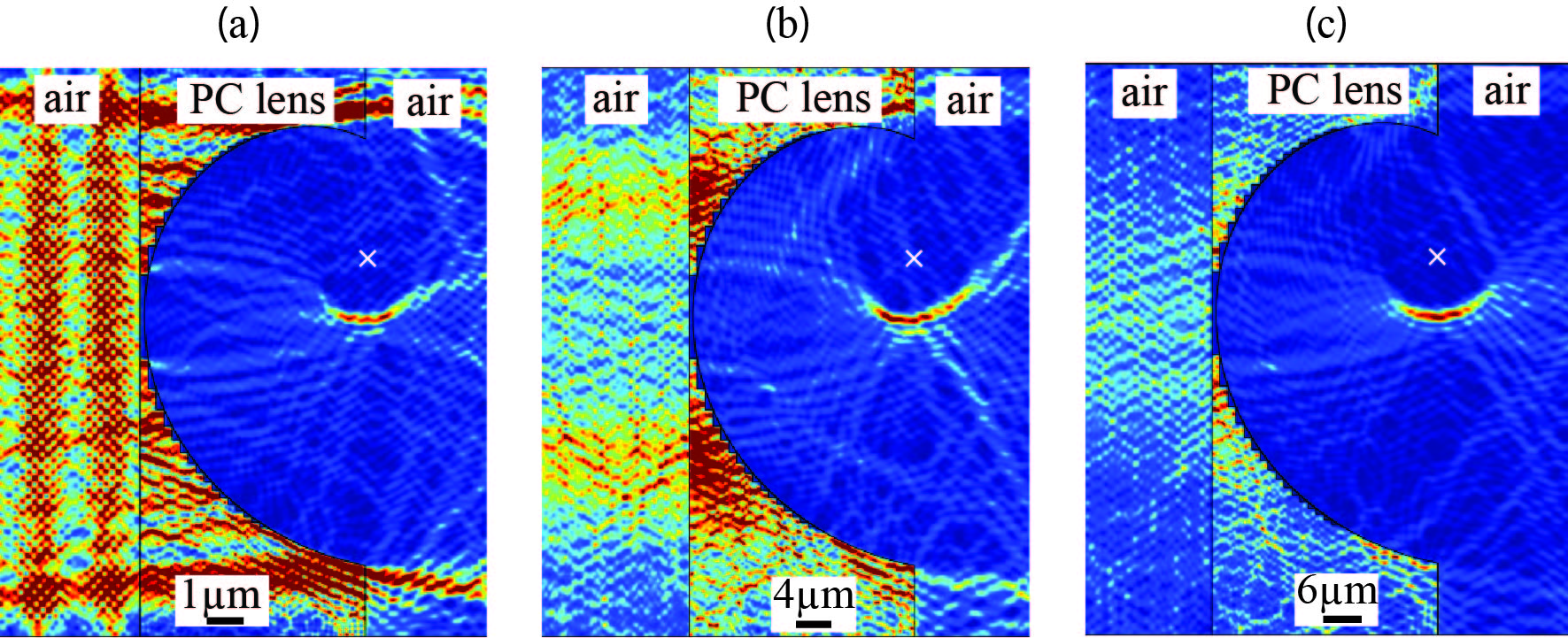}
	\caption{Generation of highly-accelerating beams using plano-concave lenses with surface-relief coating curvature given by $R_0 = 10 \lambda$ and $R' = R_0 / 2$.
		In the FEM-based numerical simulations, the time-averaged energy density at the focal region is evaluated by using ENZ materials: (a) Ag at $\lambda = 326$~nm ($\mathrm{Im}(\epsilon) = 0.7$), (b) Al:ZnO at $\lambda = 1355$~nm ($\mathrm{Im}(\epsilon) = 0.31$), and (c) n-CdO at $\lambda = 1870$~nm ($\mathrm{Im}(\epsilon) = 0.127$).}
	\label{fig06}
\end{figure*}

For the sake of illustration, we first consider a FEM-based numerical simulation showing the scattered field of a surface-relief ENZ metacoating containing 32 elements, which is deposited on the terraced exit surface of a dielectric plano-concave lens.
If the radius of the device is $R_0 = 10 \lambda$, the index of refraction of the glass substrate is $n = 1.5$, and the ENZ material is silver (we use a wavelength $\lambda= 326$~nm where Im$(\epsilon)=0.7$), the resultant time-averaged energy density of the focused wave field is shown in Fig.~\ref{fig05}(a).
Therefore, each prismatic element of the ENZ silver assembly has a width of only 217~nm.
We point out that a higher number of pieces can be used to completely cloak the terraced-concave surface of the lens, but in practical term we may restrict such number with a negligible loss of electromagnet energy due to the inherent apodization effect of high-numerical aperture geometries, as shown above.
In the numerical simulation we observe the formation of a focused field with a FWHM of $198$~nm in the geometrical focal plane and $512$~nm along the optical axis, demonstrating a near-optimal resolution performance.
We point out that a moderate focal shift is observed with an increased energy density of $53\%$ and a reduced transverse FWHM of $175$~nm.
Furthermore, the focal energy density is multiplied by a factor of $\sim 19$ when it is compared with the focusing arrangement composed of a silver-only plano-concave lens.
Surprisingly, we also compared the energy efficiency by numerically disregarding the metal losses of the silver-only plano-concave lens; in such a case the energy density is multiplied by a factor of $\sim 1.9$.
Therefore, our proposal provides a higher energy efficiency than a lossless plano-concave lens.
Finally, as occurs above, our result obtained for a TM-polarized incident plane wave is notable better than assuming a TE-polarized impinging field (not shown in the figure).

To further inspect the behavior of the coated microlenses we design the ENZ elements to form a nonparaxial accelerating beam. 
Again, the width of these elements are $\lambda / n$ and the vertical entrance surface enables the incident TM-polarized plane wave to perfectly couple to the silver assembly.
In this case, we modulate the exit surface following the equation given in (\ref{eq01}), provided that $R' = R_0 / 10$, and again $R_0 = 10 \lambda$.
Figure~\ref{fig05}(b) clearly shows the acceleration of the beam around the central point $C$, contrarily to what occurs for full-ENZ plano-concave lenses (see Fig.~\ref{fig03}(b)).

In Fig.~\ref{fig06}(a) we show the focal distribution of the time-averaged energy density of the scattered field for an increased spatial acceleration.
Now the FEM-based numerical simulation is carried out for a silver assembly which is patterned by following a radial modulation $R' = R_0 / 2$.
Our results demonstrate the feasibility of the procedure in good agreement with those obtained with a lossless ENZ microlens, previously shown in Fig.~\ref{fig02}(d).
Furthermore, such an strategy can be implemented by using other sort of ENZ materials.
Let us now consider Al:ZnO, which is a doped semiconductor whose permittivity has a vanishing real part at $\lambda = 1355$~nm; in this case $\mathrm{Im}(\epsilon) = 0.31$ \cite{Caldwell15}.
When the modulation of the radius of curvature is again given by $R_0 = 10 \lambda$ and $R' = R_0/2$, keeping as $n = 1.5$ the index of refraction of the dielectric thick lens, the resultant accelerating beam is represented in Fig.~\ref{fig06}(b).
Roughly speaking, the beam shaping is mimetically reproduced as expected from the scaling properties of the Maxwell equations \cite{Joannopoulos08}.
To complete our numerical analysis, we finally operate with the ENZ n-CdO occurring at a wavelength of $\lambda = 1870$~nm, where in addition $\mathrm{Im}(\epsilon) = 0.127$ \cite{Caldwell15}.
Due to the reduced Ohmic losses of the doped semiconductor, the enhanced quality of the nonparaxial accelerating beam is evident.

\begin{figure}[tb]
	\centering
	\includegraphics[width=.8\linewidth]{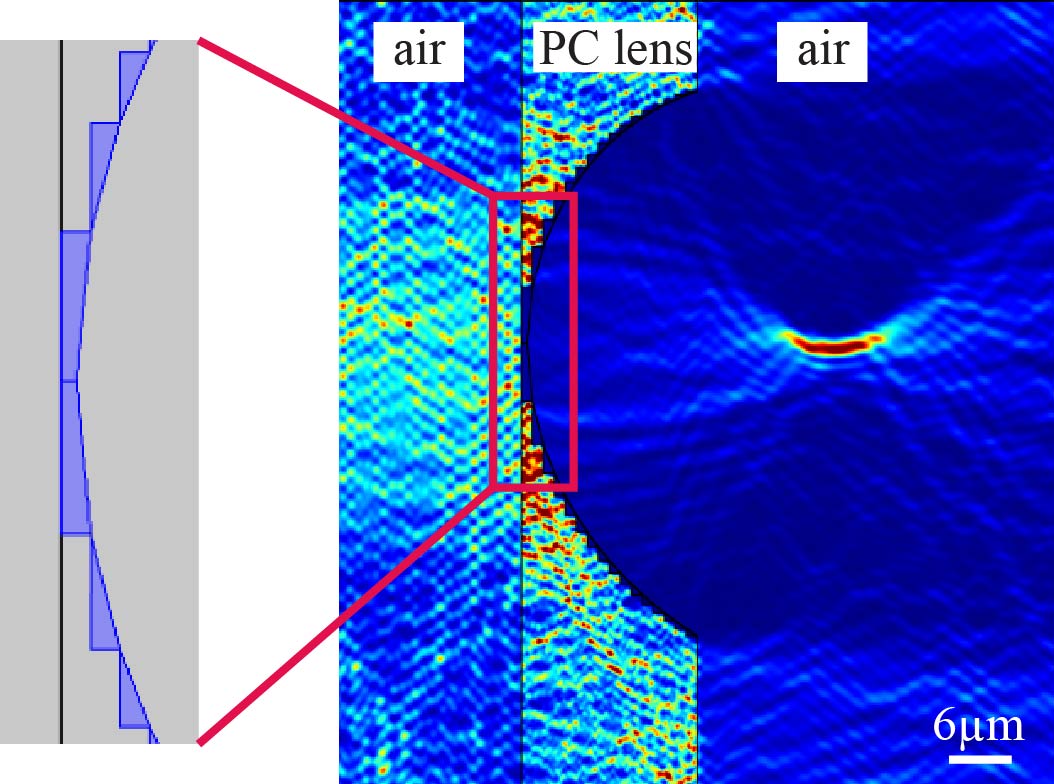}
	\caption{Spatial distribution of the time-averaged energy density generated by a plano-convace lens coated by n-CdO microprisms (as zoomed-in in the inset) at a wavelength $\lambda = 1870$~nm.}
	\label{fig07}
\end{figure}

The robustness of our procedure to focus and accelerate a plane wave within an area of a few $\mu$m$^2$ will be demonstrated next.
For that purpose, we substitute each ENZ curved element of the metacoating by a rectangular prism, in such a way that the base again has a length of $\lambda / n$, as well as the lateral face where the plane beam impinges on also remains unchanged.
On the contrary, the curved side where the beam exits from the coating is now turned into a straight hypotenuse, as depicted in the inset of Fig.~\ref{fig07}.
For the sake of illustration, once more we consider the plano-concave lens coated by n-CdO, which becomes ENZ at a wavelength of $\lambda = 1870$~nm.
As a result, the divided wavefront will be individually bent in each ENZ element with a different angle, thus constructively interfering in the focal region along the caustic curve.
However, such caustic curve is now a regular polygon approaching an incomplete circle, which leads to a fractional Talbot effect \cite{Zhang16b}.
Note that in the numerical simulations shown in Fig.~\ref{fig07}, the plano-concave dielectric lens is marginally thinner in comparison with that shown in Fig.~\ref{fig06}(c), in such a way that the geometrical numerical aperture is slightly reduced in order to account for the ENZ elements making the terraced-concave surface optically active.
Again, by comparing the distribution of the time-averaged energy density in the focal region of the original architecture shown in Fig.~\ref{fig06}(c) and the simplified model depicted in Fig.~\ref{fig07} using prismatic elements, it is evident that the pattern differences on the generated nonparaxial accelerating beams are negligible.

\section{Conclusions}

In summary, we numerically investigate the efficient generation of nonparaxial focused wave fields and accelerating beams with Bessel signature using ENZ materials.
ENZ-enabled beam shaping is inherently based on a surface effect and therefore bulk lossy devices are here substituted by transparent materials which in addition include an ENZ subwavelength coating.
We demonstrate that the locally steering of the beam can be straightforwardly executed by an assembly of subwavelength ENZ prisms with tailored hypotenuse surface, suitably deposited on the terraced-concave surface of a high-numerical-aperture lens made of a lossless dielectric.
In practical terms, the curved surface of the ENZ subwavelength prisms constituting the surface-relief coating of the plano-concave lens can be flattened for manufacturing purposes, still providing satisfactory results.
In this study we used noble metals and doped semiconductors as ENZ materials in the UV, visible and near-IR, but artificial ENZ metamaterials can be used to extend the applicability of the proposed arrangement in a broader electromagnetic spectrum \cite{Subramania12,Maas13,Gao13}.
We point out that our proposal provides an energy efficiency which is higher than that a lossless all-ENZ plano-concave lens can offer.
As concluding remark, our procedure can also be used to generate Airy beams in the paraxial regime \cite{Siviloglou07b}, accelerating nonparaxial beams with Bessel signature \cite{Kaminer12}, and even nondiffracting Bessel beam with asymmetric transverse profile \cite{Kotlyar14b}.

%To generate accelerating beams, we can use metallic gratings to artificially create extremely-high effective index of refraction \cite{Naserpour15,Naserpour15b}.
%Such technique can be used for creating focused waves with high numerical apertures \cite{Naserpour15d}.
%Even we can create light capsules by using symmetric metal-dielectric nanostructures \cite{Naserpour15c}.

\section*{Acknowledgments}

This work was supported by the Spanish Ministry of Economy and Competitiveness under contract TEC2014-53727-C2-1-R.

\bibliographystyle{elsarticle-num}

\begin{thebibliography}{10}
	\expandafter\ifx\csname url\endcsname\relax
	\def\url#1{\texttt{#1}}\fi
	\expandafter\ifx\csname urlprefix\endcsname\relax\def\urlprefix{URL }\fi
	\expandafter\ifx\csname href\endcsname\relax
	\def\href#1#2{#2} \def\path#1{#1}\fi
	
	\bibitem{Verslegers09}
	L.~Verslegers, P.~B. Catrysse, Z.~Yu, S.~Fan, Planar metallic nanoscale slit
	lenses for angle compensation, Appl. Phys. Lett. 95 (2009) 071112.
	
	\bibitem{Ishii11}
	S.~Ishii, A.~V. Kildishev, V.~M. Shalaev, K.-P. Chen, V.~P. Drachev, Metal
	nanoslit lenses with polarization-selective design, Opt. Lett. 36 (2011)
	451--453.
	
	\bibitem{Naserpour16}
	M.~Naserpour, C.~J. Zapata-Rodr{\'\i}guez, C.~D{\'\i}az-Avi{\~n}{\'o},
	M.~Hashemi, Metacoatings for wavelength-scale, high-numerical-aperture
	plano--concave focusing lenses, J. Opt. Soc. Am. B 33~(10) (2016) 2120--2128.
	
	\bibitem{Lin10}
	L.~Lin, X.~M. Goh, L.~P. McGuinness, A.~Roberts, Plasmonic lenses formed by
	two-dimensional nanometric cross-shaped aperture arrays for {Fresnel}-region
	focusing, Nano Lett. 10~(5) (2010) 1936--1940.
	
	\bibitem{Matsui12}
	T.~Matsui, H.~T. Miyazaki, A.~Miura, T.~Nomura, H.~Fujikawa, K.~Sato, N.~Ikeda,
	D.~Tsuya, M.~Ochiai, Y.~Sugimoto, et~al., Transmission phase control by
	stacked metal-dielectric hole array with two-dimensional geometric design,
	Opt. Express 20~(14) (2012) 16092--16103.
	
	\bibitem{Ishii13}
	S.~Ishii, V.~M. Shalaev, A.~V. Kildishev, Holey-metal lenses: Sieving single
	modes with proper phases, Nano Lett. 13 (2013) 159--163.
	
	\bibitem{Wang15b}
	Q.~Wang, X.~Zhang, Y.~Xu, Z.~Tian, J.~Gu, W.~Yue, S.~Zhang, J.~Han, W.~Zhang, A
	broadband metasurface-based terahertz flat-lens array, Adv. Opt. Mater. 3~(6)
	(2015) 779--785.
	
	\bibitem{Hashemi16}
	M.~Hashemi, A.~Moazami, M.~Naserpour, C.~J. Zapata-Rodr\'{\i}guez, A broadband
	multifocal metalens in the terahertz frequency range, Opt. Commun. 370 (2016)
	306--310.
	
	\bibitem{Aieta12b}
	F.~Aieta, P.~Genevet, M.~A. Kats, N.~Yu, R.~Blanchard, Z.~Gaburro, F.~Capasso,
	Aberration-free ultrathin flat lenses and axicons at telecom wavelengths
	based on plasmonic metasurfaces, Nano Lett. 12 (2012) 4932--4936.
	
	\bibitem{Yu13}
	N.~Yu, P.~Genevet, F.~Aieta, M.~A. Kats, R.~Blanchard, G.~Aoust, J.-P.
	Tetienne, Z.~Gaburro, F.~Capasso, Flat optics: Controlling wavefronts with
	optical antenna metasurfaces, IEEE J. Sel. Top. Quantum Electron. 19 (2013)
	4700423--4700423.
	
	\bibitem{Genevet15}
	P.~Genevet, F.~Capasso, Holographic optical metasurfaces: a review of current
	progress, Rep. Prog. Phys. 78~(2) (2015) 024401.
	
	\bibitem{Alu07}
	A.~Alu, M.~G. Silveirinha, A.~Salandrino, N.~Engheta, Epsilon-near-zero
	metamaterials and electromagnetic sources: Tailoring the radiation phase
	pattern, Phys. Rev. B 75 (2007) 155410.
	
	\bibitem{Navarro12}
	M.~Navarro-C\'{\i}a, M.~Beruete, M.~Sorolla, N.~Engheta, Lensing system and
	{Fourier} transformation using epsilon-near-zero metamaterials, Phys. Rev. B
	86, (2012) 165130.
	
	\bibitem{Memarian15}
	M.~Memarian, G.~V. Eleftheriades, Analysis of anisotropic epsilon-near-zero
	hetero-junction lens for concentration and beam splitting, Opt. Lett. 40~(6)
	(2015) 1010--1013.
	
	\bibitem{Pacheco14}
	V.~Pacheco-Pe{\~n}a, V.~Torres, B.~Orazbayev, M.~Beruete, M.~Navarro-C{\'\i}a,
	M.~Sorolla, N.~Engheta, Mechanical 144 ghz beam steering with all-metallic
	epsilon-near-zero lens antenna, Appl. Phys. Lett. 105~(24) (2014) 243503.
	
	\bibitem{Pacheco16}
	V.~Pacheco-Pe{\~n}a, M.~Navarro-C{\'\i}a, M.~Beruete, Epsilon-near-zero
	metalenses operating in the visible, Opt. Laser Technol. 80 (2016) 162--168.
	
	\bibitem{Caldwell15}
	J.~D. Caldwell, L.~Lindsay, V.~Giannini, I.~Vurgaftman, T.~L. Reinecke, S.~A.
	Maier, O.~J. Glembocki, Low-loss, infrared and terahertz nanophotonics using
	surface phonon polaritons, Nanophotonics 4~(1) (2015) 44--68.
	
	\bibitem{Casse08}
	B.~D.~F. Casse, W.~T. Lu, Y.~J. Huang, S.~Sridhar, Nano-optical microlens with
	ultrashort focal length using negative refraction, Appl. Phys. Lett. 93
	(2008) 053111.
	
	\bibitem{He16}
	X.~He, Z.~Huang, M.~Chang, S.~Xu, F.~Zhao, S.~Deng, J.~She, J.~Dong,
	Realization of zero-refractive-index lens with ultralow spherical aberration,
	ACS Photonics\href {http://dx.doi.org/10.1021/acsphotonics.6b00714}
	{\path{doi:10.1021/acsphotonics.6b00714}}.
	
	\bibitem{Pacheco13}
	V.~Pacheco-Pe{\~n}a, B.~Orazbayev, V.~Torres, M.~Beruete, M.~Navarro-C{\'\i}a,
	Ultra-compact planoconcave zoned metallic lens based on the fishnet
	metamaterial, Appl. Phys. Lett. 103~(18) (2013) 183507.
	
	\bibitem{Orazbayev15}
	B.~Orazbayev, V.~Pacheco-Pe{\~n}a, M.~Beruete, M.~Navarro-C{\'\i}a, Exploiting
	the dispersion of the double-negative-index fishnet metamaterial to create a
	broadband low-profile metallic lens, Opt. Express 23~(7) (2015) 8555--8564.
	
	\bibitem{Zapata14c}
	C.~J. Zapata-Rodr\'{\i}guez, M.~Naserpour, Nonparaxial shape-preserving {Airy}
	beams with {Bessel} signature, Opt. Lett. 39 (2014) 2507--2510.
	
	\bibitem{Zapata07}
	C.~J. Zapata-Rodr\'{\i}guez, Debye representation of dispersive focused waves,
	J. Opt. Soc. Am. A 24 (2007) 675--686.
	
	\bibitem{Naserpour15b}
	M.~Naserpour, C.~J. Zapata-Rodr\'{\i}guez, A.~Zakery,
	C.~D\'{\i}az-Avi{\~n}\'{o}, J.~J. Miret, Accelerating wide-angle converging
	waves in the near field, J. Opt. 17 (2015) 015602.
	\newblock \href
	{http://dx.doi.org/http://dx.doi.org/10.1088/2040-8978/17/1/015602}
	{\path{doi:http://dx.doi.org/10.1088/2040-8978/17/1/015602}}.
	
	\bibitem{Naserpour15d}
	M.~Naserpour, C.~J. Zapata-Rodr\'{\i}guez, C.~D\'{\i}az-Avi{\~n}\'{o},
	M.~Hashemi, J.~J. Miret, Ultrathin high-index metasurfaces for shaping
	focused beams, Appl. Opt. 54 (2015) 7586--7591.
	\newblock \href {http://dx.doi.org/http://dx.doi.org/10.1364/AO.54.007586}
	{\path{doi:http://dx.doi.org/10.1364/AO.54.007586}}.
	
	\bibitem{Zapata00}
	C.~J. Zapata-Rodr\'{\i}guez, P.~Andr\'es, M.~Mart\'{\i}nez-Corral,
	L.~Mu{\~n}oz-Escriv\'a, Gaussian imaging transformation for the paraxial
	{D}ebye formulation of the focal region in a low-{F}resnel-number optical
	system, J. Opt. Soc. Am. A 17 (2000) 1185--1191.
	
	\bibitem{Joannopoulos08}
	J.~D. Joannopoulos, S.~G. Johnson, J.~N. Winn, R.~D. Meade, Photonic crystals.
	Molding the flow of light, Princeton University Press, 2008.
	
	\bibitem{Zhang16b}
	Y.~Zhang, H.~Zhong, M.~R. Beli{\'c}, C.~Li, Z.~Zhang, F.~Wen, Y.~Zhang,
	M.~Xiao, Fractional nonparaxial accelerating talbot effect, Opt. Lett.
	41~(14) (2016) 3273--3276.
	
	\bibitem{Subramania12}
	G.~Subramania, A.~Fischer, T.~Luk, Optical properties of metal-dielectric based
	epsilon near zero metamaterials, Appl. Phys. Lett. 101~(24) (2012) 241107.
	
	\bibitem{Maas13}
	R.~Maas, J.~Parsons, N.~Engheta, A.~Polman, Experimental realization of an
	epsilon-near-zero metamaterial at visible wavelengths, Nat. Photonics 7~(11)
	(2013) 907--912.
	
	\bibitem{Gao13}
	J.~Gao, L.~Sun, H.~Deng, C.~J. Mathai, S.~Gangopadhyay, X.~Yang, Experimental
	realization of epsilon-near-zero metamaterial slabs with metal-dielectric
	multilayers, Appl. Phys. Lett. 103~(5) (2013) 051111.
	
	\bibitem{Siviloglou07b}
	G.~A. Siviloglou, J.~Broky, A.~Dogariu, D.~N. Christodoulides, Observation of
	accelerating {Airy} beams, Phys. Rev. Lett. 99 (2007) 213901.
	\newblock \href {http://dx.doi.org/10.1103/PhysRevLett.99.213901}
	{\path{doi:10.1103/PhysRevLett.99.213901}}.
	
	\bibitem{Kaminer12}
	I.~Kaminer, R.~Bekenstein, J.~Nemirovsky, M.~Segev, Nondiffracting accelerating
	wave packets of {Maxwell}'s equations, Phys. Rev. Lett. 108 (2012) 163901.
	
	\bibitem{Kotlyar14b}
	V.~V. Kotlyar, A.~A. Kovalev, V.~A. Soifer, Asymmetric {Bessel} modes, Opt.
	Lett. 39 (2014) 2395--2398.
	
\end{thebibliography}

\end{document}